\documentclass[preprint,12pt]{elsarticle}

\usepackage{graphics}
\usepackage[percent]{overpic}

\journal{Physica B}

\begin{document}

\begin{frontmatter}

\title{Identification of triangular-shaped defects often appeared in hard-sphere crystals grown on a square pattern under gravity by Monte Carlo simulations}

\author{Atsushi Mori$^{\ast}$ and Yoshihisa Suzuki}

\address{Institute of Technology and Science, The University of Tokushima, 2-1 Minamijosanjima, Tokushima 770-8506, Japan \\[2ex]
$^\ast$Corresponding author; \\
E-mail: atsushimori@tokushima-u.ac.jp, \\
phone: +81-88-656-9417, fax: +81-88-656-9435}

\date{manuscript received 22 April, 2014;}

%\maketitle

\begin{abstract}
In this paper, we have successfully identified the triangular-shaped defect structures with stacking fault tetrahedra.
These structure often appeared in hard-sphere (HS) crystals grown on a square pattern under gravity. 
We have, so far, performed Monte Carlo simulations of the HS crystals under gravity.
Single stacking faults as observed previously in the HS crystals grown on a flat wall were not seen in the case of square template.
Instead, defect structures with triangular appearance in $xz$- and $yz$- projections were appreciable.
We have identified them by looking layer by layer.
Those structures are surrounded by stacking faults along face-centered cubic (fcc) \{111\}.
Also, we see isolated vacancies and vacancy-interstitial pairs, and we have found octahedral structures surrounded by stacking faults along fcc \{111\}.\\

\noindent
PACS: 82.70.Dd, 61.72.Nn, 61.72.Cc, 61.50.Ks
%61.50.-f 	Structure of bulk crystals
%61.50.Ks 	Crystallographic aspects of phase transformations; pressure effects
%
%61.72.-y 	Defects and impurities in crystals; microstructure
%61.72.Cc 	Kinetics of defect formation and annealing
%61.72.Nn 	Stacking faults and other planar or extended defects
%
%81.40.-z 	Treatment of materials and its effects on microstructure, nanostructure, and properties
%81.40.Ef 	Cold working, work hardening; annealing, post-deformation annealing, quenching, tempering recovery, and crystallization
%
%82.70.-y 	Disperse systems; complex fluids
%82.70.Dd 	Colloids
\end{abstract}


\end{frontmatter}

\newpage

\section{\label{sec:intro}Introduction}

In 1997, Zhu~\textit{et~al.} discovered an effect of gravity that reduces the stacking disorder in hard-sphere (HS) colloidal crystals \cite{Zhu1997}.
The mechanism of reduction of the stacking defects due to gravity has not been understood for a long time; even if the stacking disorders occur, the particle number density is not affected.
We demonstrated the reduction of stacking disorders in the HS crystals under gravity by Monte Carlo (MC) simulations \cite{Mori2006JCP}.
By close looks at the snapshots of these simulations, we found a glide mechanism of a Shockley partial dislocation for shrink of an intrinsic stacking fault for the HS crystals grown in face-centered cubic (fcc) $\langle$001$\rangle$ \cite{Mori2007}.
A key is fcc $\langle$001$\rangle$ stacking; in this stacking the stacking fault runs along oblique \{111\} and the intrinsic stacking fault is terminated at its lower end by the Shockley partial dislocation.
One can understand buoyancy of the Shockley partial dislocation because its dislocation core accompanies a particle deficiency of 1/3 lattice line.
Also, the elastic field  due to the Shockley partial dislocation yields a driving force for the Shockley partial dislocation to move toward shrinking of the intrinsic stacking fault \cite{Mori2009}.
That is, the gravity as well as the elastic field gives a driving force for shrinking of the stacking faults.
Additionally, we calculated the cross term of elastic fields due to gravity and the Shockley partial dislocation~\cite{Mori2010}.
This term also gives a driving force for shrinking of the staking faults.

In 1993, Biben~\textit{et~al.} studied the colloidal crystals under gravity relying on a density functional theory with MC simulations of the HS and screened Coulomb systems \cite{Biben1993}.
The crystal defect was not their subject; their interest was in computation of density profiles on the basis of macroscopic equilibrium condition and their relation to the interparticle interaction.
The number of particle on a unit area of the bottom wall (hereafter $n_s^*$) was enough large.
 Thus, the onset of the crystal defect could be seen in a vertical density profile; oscillatory amplitude decreased non-monotonically with the altitude for $n_s^*$ = 40 and $g^*$ = 0.4 (the dimensionless quantity $g^*$is defined as $g^*$ = $mg\sigma/k_BT$, which is an indicator of the strength of the gravity relative to the thermal energy, where $m$ is the (buoyant) mass, $g$ the acceleration due to gravity, $\sigma$ the HS diameter, $k_B$ Boltzmann's constant, and $T$ the temperature).

In 1989, Pusey~\textit{et~al.} found formation of mixture of fcc and random hexagonal close pack (rhcp), which is random stacking of fcc \{111\} hexagonal planes, in sediments of an HS colloid under normal gravity by  using a laser light scattering \cite{Pusey1989}.
Viewing fcc crystals along $\langle$111$\rangle$, the stacking is of ABC type, where A, B, and C stand for the three types of the hexagonal planes on the basis of the lateral position of particles.
On the other hand, for hexagonal close-pack (hcp) the stacking sequence is of ABAB type.
The rhcp corresponds to the random sequences of A, B, and C.
Zhu~\textit{et~al.} found that the rhcp forms for the HS colloid in microgravity \cite{Zhu1997} while under normal gravity mixture of rhcp and fcc forms as reported in Ref.~\cite{Pusey1989}.

A criticism can arise that the situation in Ref.~\cite{Mori2007} is unrealizable.
That is, fcc $\langle$001$\rangle$ stacking there was forced by a small simulation box with the periodic boundary condition (PBC).
Against this criticism, we have emphasized that a stress possessing the same symmetry can be realized by using a patterned substrate (template) with the same symmetry.
Usage of the template in the colloidal crystallization was proposed by van Blaaderen~\textit{et~al.} \cite{vanBlaaderen1997} (the colloidal epitaxy).
Recently, we have successfully replaced the driving force for fcc $\langle$001$\rangle$ stacking (the stress from a small PBC simulation box) with a stress from a square pattern on the bottom wall~\cite{Mori2011WJE,Mori2011JCG,Mori2012WJE}.
We note that the colloidal epitaxy on a square pattern was already demonstrated by Lin~\textit{et~al.} \cite{Lin2000}.
An advantage of the use of square pattern as compared to fcc (001) pattern is that the lattice constant of the crystal grown on the template can, in principle, be adjusted by matching the crystal on the lattice lines instead of the lattice points.

In this way, we have successfully demonstrated the reduction of stacking disorders in the colloidal epitaxy under gravity, and given a reasonable understanding of the mechanism.
The reduction of defects has, however, not been perfect.
Defect structures with triangular appearance in $xz$- and $yz$- projections have sometimes been appreciable.
The main purpose of this paper is to identify those triangular defect structures in a framework of geometrical crystallographic consideration.
The defect structures with triangular appearance were already seen in simulations of a small system size \cite{Mori2011WJE}.
We could, however, not rule out the system size effect.
That is, the triangular shapes might be resulted so that stress from the small simulation box be minimized.
Indeed, triangular shapes were not isolated with each other.
By doubling the horizontal system size, isolated triangular shapes have formed \cite{Mori2012WJE}.
In this paper, we will take close looks at these defect structures.

It is helpful for understanding the significance of the present study to distinguish the present defect structures from the similar ones.
The triangular-shaped defect structures were observed by Meijer~\textit{et~al.} \cite{Meijer2007}, de Villeneuve~\textit{et~al.} \cite{Villeneuve2007}, and Hilhorst~\textit{et~al.} \cite{Hilhorst2010}.
Meijer~\textit{et~al.}'s observations were, however, in a hexagonal layer (though de Villeneuve~\textit{et~al.}'s aim was observations of grains, these were done in the hexagonal layers).
That is, triangular-shaped islands formed in an fcc \{111\} layer in fcc \{111\} stacking --- note that in the present study we investigate fcc (001) stacking.
They made three-dimensional (3D) analysis and the islands observed were three-dimensionally extended; however, 3D outer shapes were not put in relieves.
The triangular shapes there were purely originated in the difference in lateral stacking position as compared to the surroundings.
Meijer~\textit{et~al.} made a picture of a 3D dislocation network; however, it was unfortunate that no 3D structures formed by stacking disorders were drawn.
Also, Hilhorst~\textit{et~al.} showed 2D confocal images, in which the triangular-shaped structures in a horizontal hexagonal layer were seen.
The origin of the observed triangular shapes there were in the difference in the lateral positions of the particles in the hexagonal layers, too.
Besides, 3D consideration was done.
That is, slanted stacking faults were observed; however, no 3D structure formed by stacking disorders was drawn, too.
The same situation has been seen in a recent simulation study \cite{Marechal2011}.

The paper is organized as follows.
Section~\ref{sec:simulation} describes the system and simulation method.
We take layer-by-layer looks for the bottom, middle, and upper regions in Sec.~\ref{sec:result}.
Although the main purpose is to identify the triangular-shaped defect structures, we analyze other defect structures.
In particular, in Sec.~\ref{sec:bottom} the majority of defects is point defects.
On the other hand, in Sec.~\ref{sec:upper} we postpone the conclusion because the defect structure in the upper region seems to be affected by the system size.
Identification of the compact triangular-shaped defects structures is done in Sec.~\ref{sec:middle}.
We note that, as a first step, we make geometrical crystallographic considerations.
Discussion is given in Sec.~\ref{sec:discussion}.
Section~\ref{sec:conclusion} concludes this paper, along with remarks on future studies.

\section{\label{sec:simulation}System and simulation}

$N$ = 26634 HSs were confined between a flat top wall at $z$ = $L_z$ and a square template at $z$ = 0.
The PBC was imposed in the horizontal $x$ and $y$ directions.
The system size was as $L_x$ = $L_y$ = $25.09\sigma$ (thus $n_s^*$ = 42.3) and $L_z$ = $1000\sigma$.
The square pattern on the bottom wall was as follows: the groove width was $0.707106781\sigma$ and the side-to-side separation between adjacent grooves was $0.338\sigma$ (see Ref.~\cite{Mori2012CL} for a note on the significant digits); the numbers of groves along $x$- and $y$- axes were 24.
The diagonal length of the intersection of the grooves running in $x$ and $y$ directions is, thus, $0.9999999997\sigma$.
As a results, an HS did not fall into but fitted to the intersection.
An illustration of the square pattern is seen in a chapter of a book \cite{Mori2011InTech}.
We maintained the gravitational number $g^*$ constant.
On the other hand, in early days~\cite{Mori2011WJE,Mori2011JCG}, we controlled gravity stepwise as proposed in Ref.~\cite{Mori2006JCP} as a method to avoid the metastable polycrystalline state.
We continued MC simulations for $5.12\times10^7$ Monte Carlo cycles (MCC).
Here, one MCC is defined such that one MCC contains $N$ MC moves.
The maximum displacement of the MC move of a particle was fixed at $0.06\sigma$.

A note on $g^*$, which plays a central role in sedimentation, is given here.
This quantity is identical to the reduced inverse gravitational length $\sigma/l_g$.
Here, the gravitational length is defined as $l_g = k_BT/mg$, lengths over which the gravitational effects work.
$g^*$ is also equivalent to the Peclet number, except for a proportional constant, which is the ratio between the drift and diffusion.

\section{\label{sec:result}Results --- identification of defects in snapshots}

We take layer-by-layer looks at snapshots in Sections~\ref{sec:bottom}-\ref{sec:upper}.
Analyses are made only on a case of $g^*$ = 1.6 because the defect structures are more appreciable in this case.

\subsection{\label{sec:triangular}Triangular-shaped defect structures}

\begin{figure*}[htb]
\includegraphics{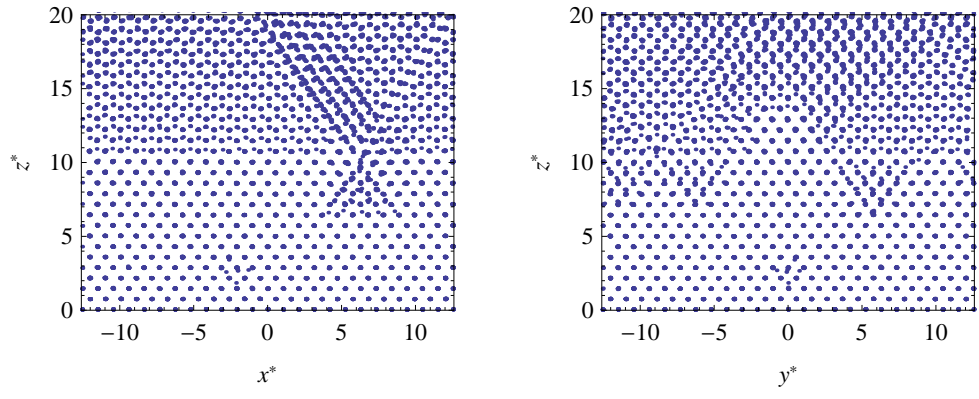}
\caption{\label{fig:snapshot}Snapshot at 4.5 $\times10^7$th MCC for $g^*$ = 1.6: $xz$- (left) and $yz$- (right) projections are shown.}
\end{figure*}

Snapshot ($xz$- and $yz$- projections) at $4.5\times10^7$th MCC for $g^*$ = 1.6 of a previous simulation~\cite{Mori2012WJE} is shown in Fig.~\ref{fig:snapshot}.
In this figure, $xz$- and $yz$- side views are depicted.
We have already shown snapshots at $1\times 10^5$th,  $1.5\times 10^6$th, and  $5\times 10^7$th MCCs \cite{Mori2012WJE}.
$1\times 10^5$th MCC was just after the instant when gravity was switched on.
At this instant, a less defective crystal in a bottom region, a defective crystal above it, and a disordered structure were already seen.
In the defective crystal, splitting of the lattice lines took place.
One can understand this as a result of intersection of the lattice lines with a stacking disorder such as a stacking fault.
If a lattice line runs across a stacking fault, the lattice line is separated into two by the Burgers vector of a Shockley partial dislocation.
At $1.5\times 10^6$th MCC, which was at the end of the rapid sink of the center of gravity, the mid disordered region was converted into a part of the less defective crystal.
That is, the separation of the lattice lines existed at $1\times 10^5$th MCC disappeared.
Remarkable differences are not seen between the snapshots at $1.5\times 10^6$th and  $5\times 10^7$th MCCs.
Just tiny narrowing of the projections of the lattice lines around $(y/\sigma, z/\sigma)$ = (10,8), for example, occurred.
Hereafter, we will use $^*$ to indicate the reduced unit such as $x^* \equiv x/\sigma$ (in figures the asterisks will be omitted sometimes to make the figures compact).
Also, Fig.~\ref{fig:snapshot} is essentially the same as snapshots already published \cite{Mori2012WJE}.
Triangular-shaped defect structures are noticeable.
In other simulations using a different series of random numbers, not only triangular-shaped defect structures but also disorders which are seemingly the same as that seen at $(x^*,y^*,z^*)$ = ($-2$, 0, 2) in Fig.~\ref{fig:snapshot} exist.
Let us close looks at the configuration at $4.5\times10^7$th MCC below.

\subsection{\label{sec:bottom}Bottom region}

\begin{figure}[htbp]
\includegraphics{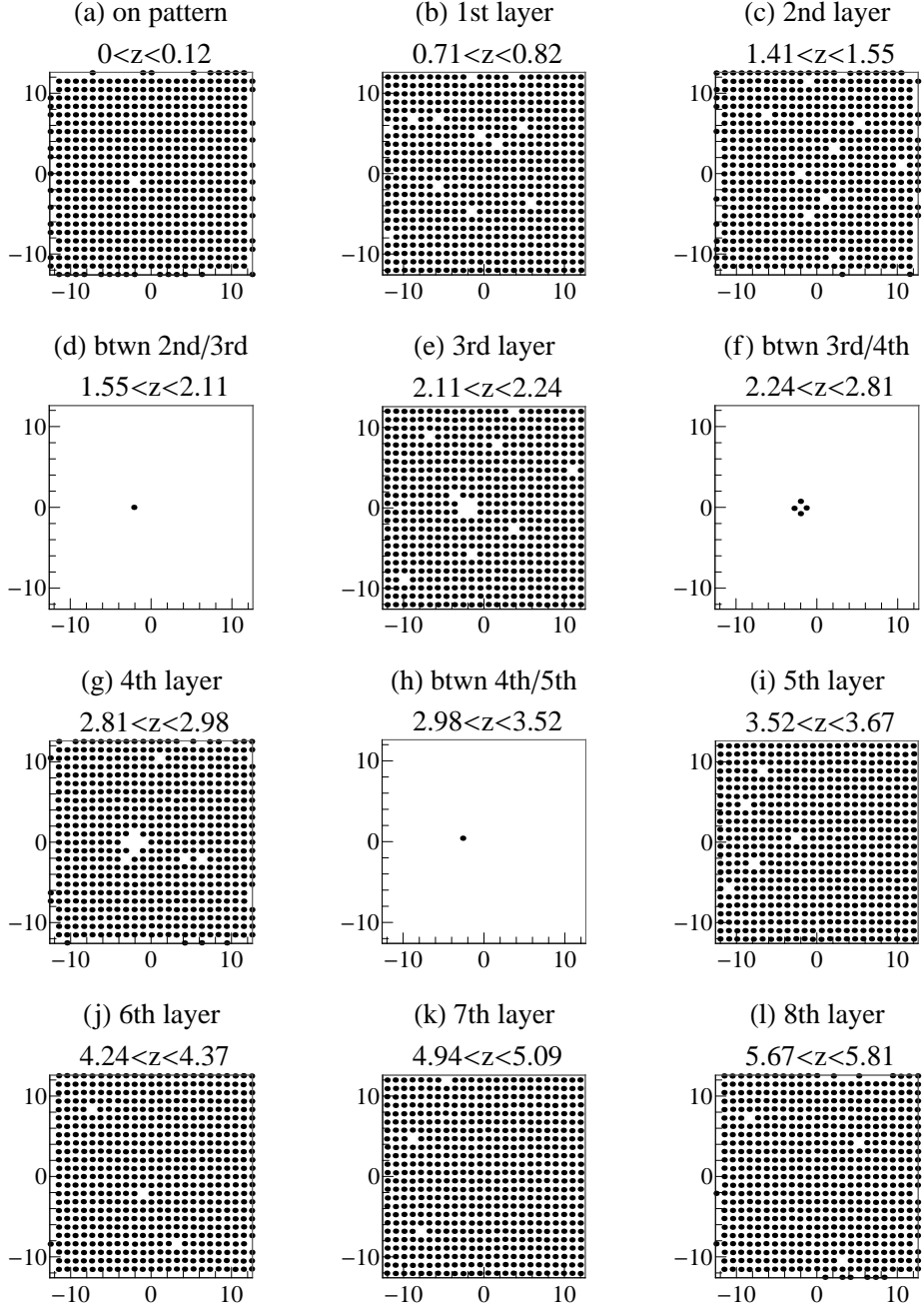}
\caption{\label{fig:bottomslice}
Top views of slices of each layer in the bottom region (0 $\leq$ $z^*$ $\leq$ 5.81) [(a-c), (e), (g), (i-l)] and those of three slices of two inter layer regions [(d), (f), and (h)].
Horizontal and vertical directions are, respectively, $x$ and $y$ directions.}
\end{figure}

We examine horizontal slices in the range $0 \leq z^* \leq 5.81$ viewed from the top (Fig.~\ref{fig:bottomslice}).
In Figs.~\ref{fig:bottomslice}, \ref{fig:middleslice}, and \ref{fig:upperslice}, slices cut parallel to $xy$-plain viewed from the top are arranged from the bottom to the top.
The thickness of each slice was not constant, but was adjusted so as for the crystal plane or interstitial region to be appropriately identified.
Also, we note that the points in those figures indicate the central positions of particles in $xy$-plain.

The slice on the bottom pattern ($0<z^*<0.12$) is shown in Fig.~\ref{fig:bottomslice}~(a).
We observe a vacant site in this view.
Also, vacant sites exist in the first layer ($0.71<z^*<0.82$) [Fig.~\ref{fig:bottomslice}~(b)].
Because there is no particle in the regions $0.12<z^*<0.71$ (between the bottom and first layers) and $0.82<z^*<1.41$ (between the first and the second layers), those vacant sites are identified with isolated vacancies.

On the other hand, looking around $(x^*,y^*)$ = $(-2,0)$ in Fig.~\ref{fig:bottomslice}~(d), (f), and (h), we confirm formation of octahedral structure of six particles.
In addition, particles on the regular lattice planes at corresponding horizontal locations are absent [Fig.~\ref{fig:bottomslice}~(c), (e), (g), (i), and (j)].
It means that the six-particle cluster is formed by simultaneous translation of those six particles downward [note that the shape of the void around $(x^*,y^*)$ = $(-2,0)$ in Fig.~\ref{fig:bottomslice}~(g) coincides to that of the cluster around the same horizontal location in Fig.~\ref{fig:bottomslice}~(f)].
In this way, we have identified the defect at $(x^*,y^*,z^*)$ $\cong$ $(-2,0,2.5)$ in Fig.~\ref{fig:snapshot}.
This defect structure can be termed as ^^ ^^ a stacking fault octahedron" because it is surrounded by stacking faults. 

Besides this defect structure, one can find only isolated vacancies.
That is, there are no interstitial particles except for those in Fig.~\ref{fig:bottomslice}~(d), (f), and (h).
Existence of isolated vacancies was reproducible; many such vacancies were seen even in a single simulation (and in the above regions as will be shown later).
Confirmation of reproducibility for the ^^ ^^ stacking fault octahedra" needs, however, more observations.
Of course, in simulations with other random numbers, octahedral structures were seen.
We wish to postpone the conclusion on the ^^ ^^ stacking fault octahedra" because the main concern is identification of the triangular-shaped defect structures.

\subsection{\label{sec:middle}Middle region}

\begin{figure}[htbp]
\includegraphics{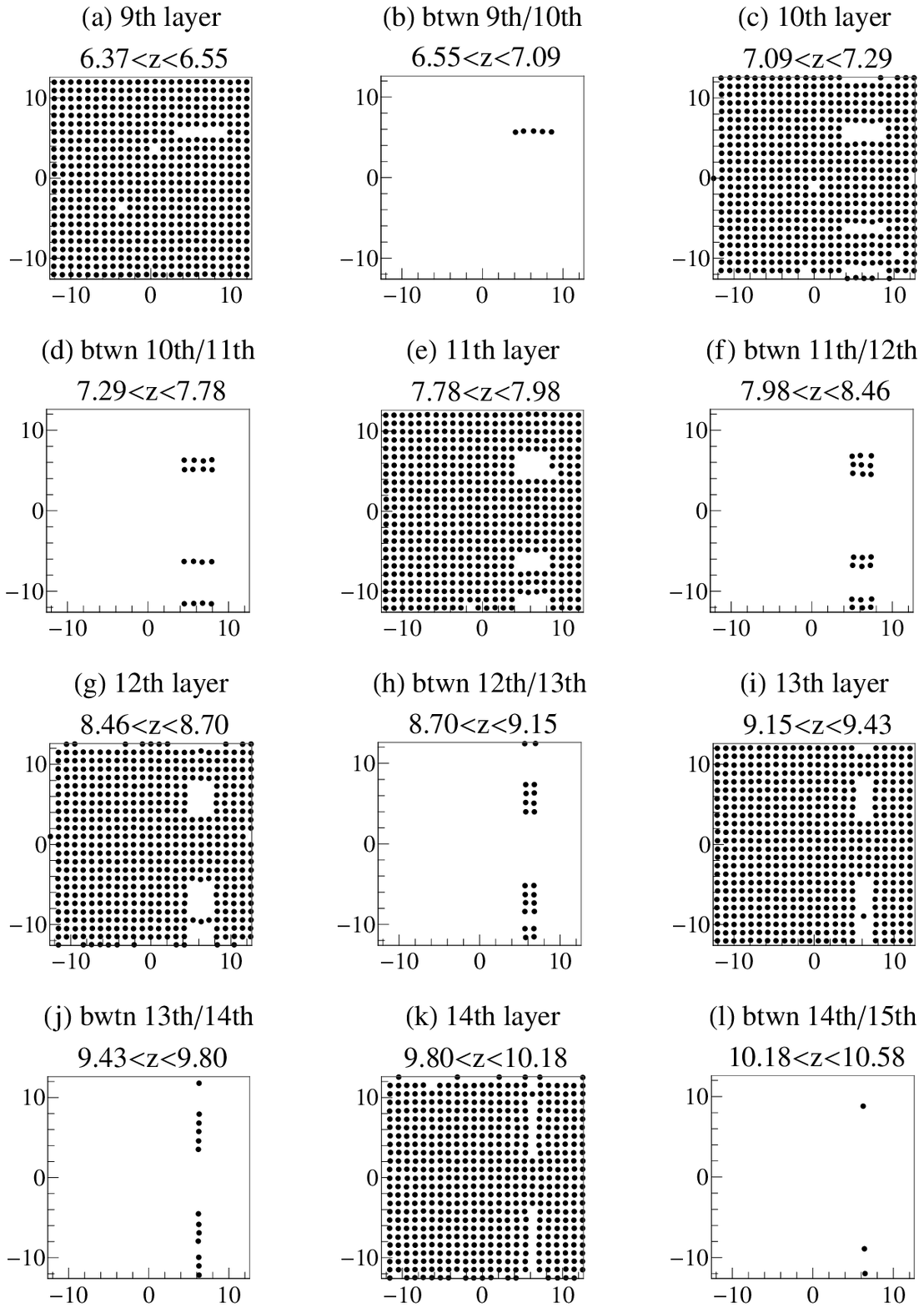}
\caption{\label{fig:middleslice}
Top views of slices of each layer in the lower region (5.81 $\leq$ $z^*$ $\leq$ 10.85) [(a), (c), (e), (g), (i), and (k)] and those of three slices of two inter layer regions [(b), (d), (f), (h), (j), and (l)].
Horizontal and vertical directions are, respectively, $x$ and $y$ directions.}
\end{figure}

We take layer-by-layer looks at horizontal slices in the region 5.81 $\leq z^*$ $\leq$ 10.85 (Fig.~\ref{fig:middleslice}), which is the region just above Fig.~\ref{fig:bottomslice}.
At first, let us confirm the isolated vacancies.
Two isolated vacancies in Fig.~\ref{fig:middleslice}~(a), four in Fig.~\ref{fig:middleslice}~(c), one in Fig.~\ref{fig:middleslice}~(e), and one in Fig.~\ref{fig:middleslice}~(k) are seen.
No interstitial particles are associated with those vacancies.
That is, as in Fig.~\ref{fig:bottomslice} those defects are isolated vacancies.

\begin{figure}[htbp]
\begin{center}
\includegraphics{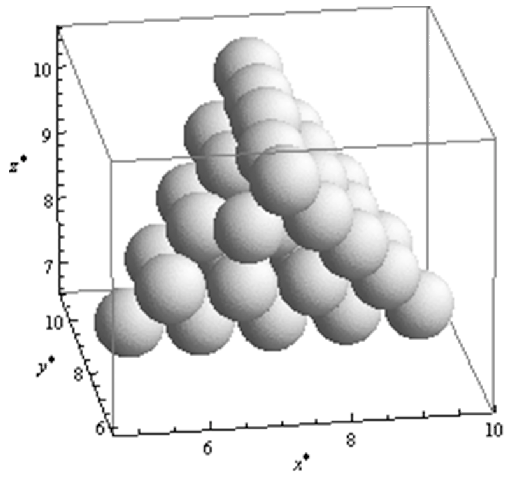}
\end{center}
\caption{\label{fig:tetraherdon}
An example of a 3D picture of a tetrahedron.
A tetrahedron at $(x^*,y^*,z^*)$ $\cong$ $(7,8,8)$ is magnified.
One can confirm that the cluster is surrounded by \{111\} and the bottom edge is along $\langle 1\bar{1}0 \rangle$ and the top edge along $\langle 110 \rangle$.
We note that due to deformation one particle on the left line on the second layer from the top and two particles on the front line on the middle layer are not displayed.}
\end{figure}

Now, we do confirm the stacking fault tetrahedra.
Let us look at the right-upper region in the slices of Fig.~\ref{fig:middleslice}~(b), (d), (f), (h), and (j) from the below to the above.
In Fig.~\ref{fig:middleslice}~(b), a line of five particles elongated in $x$ direction is seen.
In Fig.~\ref{fig:middleslice}~(d), a rectangle with four particles in $x$ direction and two particles in $y$ direction is seen.
In Fig.~\ref{fig:middleslice}~(f), a square of three by three particles is seen.
In Fig.~\ref{fig:middleslice}~(h), a rectangle with two particles in $x$ direction and four particles in $y$ direction is seen.
In Fig.~\ref{fig:middleslice}~(j), a line of five particles elongated in $y$ direction is seen.
One can understand this result as a tetrahedral cluster with an edge along $x$ direction in Fig.~\ref{fig:middleslice}~(b) and with an edge along $y$ direction in Fig.~\ref{fig:middleslice}~(j).
Taking into consideration the fact that $\langle 1\bar{1}0 \rangle$ is along $x$ direction and $\langle 110\rangle$ along $y$ direction~\cite{Mori2006STAM}, we see that the bottom edge is along $\langle 1\bar{1}0 \rangle$ and the top edge along $\langle 110 \rangle$, and the tetrahedron is surrounded by ($\bar{1}\bar{1}\bar{1}$), (11$\bar{1}$), (1$\bar{1}$1) and ($\bar{1}$11).
We note that ($\bar{1}\bar{1}\bar{1}$) and (11$\bar{1}$) intersect with each other on $\langle 1\bar{1}0 \rangle$, and (1$\bar{1}$1) and ($\bar{1}$11) on $\langle 110\rangle$.
A  3D picture of the tetrahedron is shown in Fig.~\ref{fig:tetraherdon}.
In this way, we have confirmed a stacking fault tetrahedron.
Of course, stair-rod dislocations (see Ref.~\cite{Hirth1982}) form at the edges of the tetrahedra.
The tetrahedral clusters undergo translations by the Burgers vector of this partial dislocation.
Let us take looks at other two tetrahedra in the right-bottom region in Fig.~\ref{fig:middleslice}~(d), (f), (h), and (i).
Two parallel lines of four particles elongated in $x$ direction are seen in Fig.~\ref{fig:middleslice}~(d).
Two lines of three by two particles are seen in Fig.~\ref{fig:middleslice}~(f).
Two lines of two by three particles are seen in Fig.~\ref{fig:middleslice}~(h).
Two coaxial lines of four particles elongated in $y$ direction are seen in Fig.~\ref{fig:middleslice}~(i).
In the last observation, we have taken into account the PBC in $y$ direction.
Interpretation of those observations is entirely the same as that for the tetrahedron in the right-above.
In those ways, we have identified the defect structures around ($x^*$,$y^*$,$z^*$) = $(7,8,8)$, $(7,-7,8)$, and $(7,-11,8)$ in Fig.~\ref{fig:snapshot}.
Reproducibility of the stacking fault tetrahedra was confirmed by simulations with other random numbers.

\subsection{\label{sec:upper}Upper region}

\begin{figure}[htbp]
\includegraphics{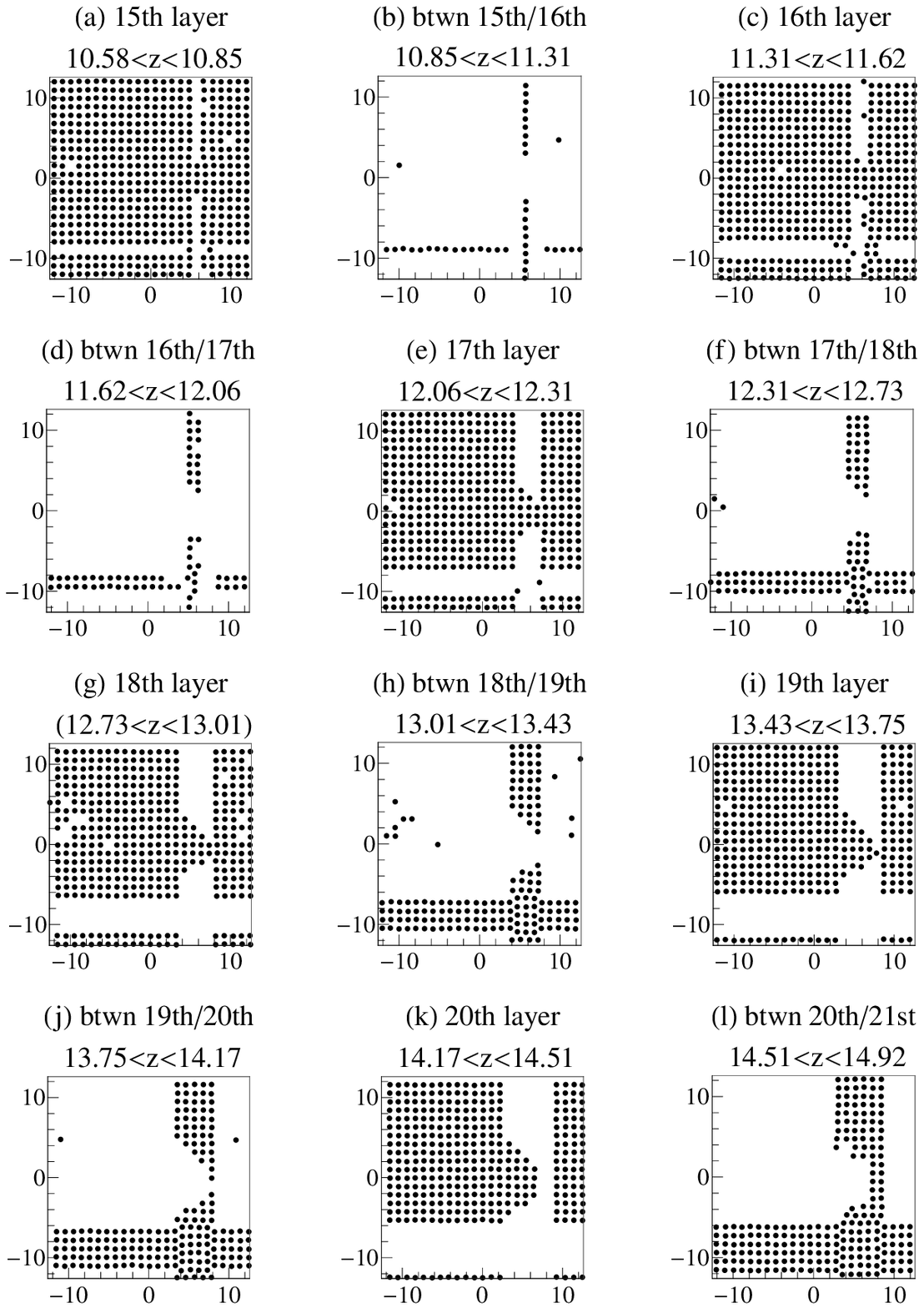}
\caption{\label{fig:upperslice}
Top views of slices of each layer in the lower region (10.85 $\leq$ $z^*$ $\leq$ 14.92) [(a), (c), (e), (g), (i), and (k)] and those of three slices of two inter layer regions [(b), (d), (f), (h), (j), and (l)].
Horizontal and vertical directions are, respectively, $x$ and $y$ directions.}
\end{figure}

We take layer-by-layer looks at horizontal slices in the region 10.84 $\leq z^*$ $\leq$ 14.92 (Fig.~\ref{fig:upperslice}); we are traversing the system from the bottom to the top.
At first, let us look at Fig.~\ref{fig:upperslice}~(a) and (b).
There are two vacancy-interstitial pairs around ($x^*$,$y^*$) = $(-10,2)$ and $(10,4)$.
Looking around $(x^*,y^*)$ = $(-11,1)$ in Fig.~\ref{fig:upperslice}~(e-i), we find a cluster of complicated form.
In the right region in Fig.~\ref{fig:upperslice}~(g) and (h), four vacancy-interstitial pairs are seen.
Vacancy-interstitial pairs are also seen at $(x^*,y^*)$ $\cong$ $(-11,5)$ and $(10,5)$ in Fig.~\ref{fig:upperslice}~(i) and (j).
Vacancy-interstitial pairs were observed in simulations with other random numbers.

Next, let us try to identify the triangular-shaped defect structure, the bottom vertex of which locates around ($x^*$,$z^*$) = $(6,11)$.
In Fig.~\ref{fig:upperslice}~(b), there is a line at $x^*$ $\cong$ 6.
Taking into consideration the PBC, this is a single line.
In Fig.~\ref{fig:upperslice}~(d), we see a line with two-particle thick around the same $x$ position.
This line contains some defects.
However, we see particles at the corresponding horizontal positions in Fig.~\ref{fig:upperslice}~(c), an intersection between a line along $x$ direction at $y^*$ $\cong$ $-9$ and that along $y$ direction at $x^*$ $\cong$ 5.
In Fig.~\ref{fig:upperslice}~(f), the line width becomes of three particles.
In Fig.~\ref{fig:upperslice}~(h), the line width becomes of four particles.
In Fig.~\ref{fig:upperslice}~(j), the line width becomes of five particles, and so on, except for the intersection at $(x^*,y^*)$ $\cong$ $(6,-9)$.
We note that this line is terminated by (100) at $(x^*,y^*)$ $\cong$ $(6,-3)$ and (010) at $(x^*,y^*)$ $\cong$ $(6,2)$.
In this way, we have reached to an understand of formation of a ^^ ^^ V-shaped" defect structure.
Now, we concentrate on the line along $x$ direction at $y^*$ $\cong$ $-9$ in Fig.~\ref{fig:upperslice}~(b).
This is a single line due to the PBC.
In Fig.~\ref{fig:upperslice}~(d), the line becomes two-particle thick.
As for the line along $y$ direction in Fig.~\ref{fig:upperslice}~(d) this line contains defects; however, there exist particles in the corresponding horizontal positions in Fig.~\ref{fig:upperslice}~(c).
The line width becomes of three particles in Fig.~\ref{fig:upperslice}~(f), becomes of four particles in Fig.~\ref{fig:upperslice}~(h), becomes of five particles in Fig.~\ref{fig:upperslice}~(j), and so on, except for the intersection at $(x^*,y^*)$ $\cong$ $(6,9)$.
In this way, V-shaped defect structure formation was revealed.
At the bottom of a V-shaped defect structure, a stair-rod dislocation forms.
Accordingly, the particles in a V shape undergo translation by the Burgers vector of the corresponding partial dislocation.
In the intersecting region of V shapes along $x$ and $y$ directions, particles undergo translations twice.
It is anticipated that these V-shaped defect structures were affected by the system size.
Thus, there is a possibility that we saw a part of an enlarged stacking fault tetraheron.
Therefore, we will not discuss more about it.
Of course, V-shaped defect structures are observed in simulations with other random numbers, sometimes with accompanied particles.

We note on the splits of the projections of the lattice lines in the region above $z^*$ $\cong$ 11 in Fig.~\ref{fig:snapshot}.
The splits seen in $xz$-projection are due to the V-shaped defect structure along $x$ direction.
Also, we see the splits in $yz$ projection, except for the region around $y^*$ = 0.
Remember that the V-shaped defect structure along $y$ direction is terminated at $(x^*,y^*)$ $\cong$ $(6,-3)$ and $(6,2)$.
The region in which the split does not occur corresponds to the region between them.

\section{\label{sec:discussion}Discussion}

\subsection{\label{sec:VI}Vacancy-interstitial pairs}

We note that no vacancy-interstitial pairs formed in the region $z^*$ $\leq$ 10.84 whereas such pairs formed in the region above.
In the lower region, the crystal is more compressed because the pressure $P(z)$ at the altitude $z$ is governed by the mechanical balance equation

\begin{equation}
\frac{\partial P}{\partial z} = -mg\rho(z).
\end{equation}
Here, $\rho(z)$ is the coarse-scale particle number density at $z$.
It is conjectured that the interstitial particles are, therefore, kicked out due to gravitational tightness in the bottom region.
On the other hand, the crystal in the upper region is less compressed.
So, the interstitial particles are allowed.
In this respect, the vacancy-interstitial pairs observed might be based on fluctuation phenomena (pair creation and annihilation) as a possibility.
Whether the vacancy-interstitial pairs remain as stable for a considerable time such as the Frenkel pairs or not can be judged by viewing movies (such as Supplementary videos of Ref.~\cite{Mori2014}).
Since the purpose of this paper is identification of the triangular-shaped defect structures, we wish to remain this as a future study.

\subsection{\label{sec:single}Single stacking faults}

We should discuss about the absence of single stacking faults, which are terminated at their lower ends by Shockley partial dislocations and whose upper ends go into a disordered fluid or a grain boundary.
As mentioned in Sec.~\ref{sec:intro}, we discovered a glide mechanism of the Shockley partial dislocation for shrinking of the intrinsic stacking fault, which runs along \{111\} oblique direction \cite{Mori2007}.
Also, we calculated driving forces for shrinking of the intrinsic stacking fault \cite{Mori2009,Mori2010}.
Therefore, it is suggested that such single stacking faults promptly disappear in an early period; the strength of gravity was not so sufficiently small that such single stacking fault remained in a late period.
Staring with a configuration including an intentional defect structure is of interest; in particular, we are interested in starting with a configuration including a stacking fault to assert this conjecture.
We observed stacking fault tetrahedra and octahedral defect structures.
From geometrical crystallographic considerations, such structure are sessile.
This may be a reason why those types of defect structures were appreciable.

\subsection{\label{sec:comparison}Comparison with other simulations and experiments}

In this paper, we have identified the stacking fault tetrahedra in the HS crystals for the first time; the system sizes of our previous simulations have, so far, not been enough large.
Although Hoover and Ree determined the condition of the crystal-fluid phase transition in the HS system in 1968~\cite{Hoover1968}, the effects of the defects were not taken into account.
Also, in the first direct crystal/fluid coexistence by molecular dynamics (MD) simulations \cite{Mori1995}, the effect of the defects was not taken into consideration.
In 1971, Bennett and Alder calculated the free energy of vacancy formation in the HS crystal at various densities by MD simulations \cite{Bennett1971}.
In 1998, Davidchack and Laird performed MD simulations of the HS crystal/fluid interface \cite{Davidchack1998}, in which they observed the vacancy propagation in the crystal.
In 2001, Pronk and Frenkel calculated the interstitial density \cite{Pronk2001}.
Effect of the point defects in the HS crystal was dealt in this way.

In turn, effect of the planar defects in the HS crystal have been focused.
Experimental studies are leading.
Besides indirect scattering studies \cite{Pusey1989,Dux1997,Kegel2000,Dolbnya2005}, direct observations of the defects in the particle level have been developed.
In 1997, Elliot~\textit{et~al.} observed stacking disorders in the HS colloidal crystals by using phase-contrast microscopy \cite{Elliot1997}.
Confocal microscopy is recently used to make real-space images of fluorescent colloidal particles.
Hoogemboom~\textit{et~al.} observed stacking faults in colloidal crystals grown by sedimentation \cite{Hoogenboom2002}.
In their study, the softness of interparticle interaction was varied from slightly repulsive to the HS like.
Stacking disorders in lateral layers of the HS colloidal crystal were observed by the confocal microscopy \cite{Villeneuve2007}.
Stacking faults have more recently been observed in the HS colloidal crystals grown by sedimentation, too \cite{Hilhorst2010}.
Stacking faults accompanying a partial dislocation were observed in colloidal crystals grown by a colloidal epitaxy by confocal microscopy \cite{Schall2004}.
MC simulations of stacking disorder in the HS crystals have been performed \cite{Villeneuve2007,Miedema2008}.
Unfortunately, their simulations were of on-lattice type; therefore, the vibrational entropy was not taken into account.
Nevertheless, the stacking disorders existed in their simulations.
It gives a strong support for existence of the stacking disorder in the HS crystals.

The present paper has a different aspect; not only the point defects and planar defects but also 3D defect structures such as stacking fault tetrahedra and ^^ ^^ stacking fault octahedra" are concerned.
By entropic consideration one knows that point defects such as vacancies, interstitials, and the Frenkel pairs are necessarily introduced in crystals at a non-zero temperature.
Also, at a non-zero temperature the entropic term associated with the stacking disorders may overcome the ^^ ^^ bonding" energy due to the stacking disorder. 
On the other hand, because the elastic energy due to a dislocation increases with $R$ (as $\propto \ln R$ \cite{Hirth1982}), the single (isolated) dislocations cannot exist in an infinite crystal in equilibrium.
Here, $R$ is the dimension over which the elastic field expands.
We have found octahedral defect structures and stacking fault tetrahedra; elastic fields for those defects are compacted such as for a dislocation dipole.

We note that whereas the stacking fault tetrahedra are commonly seen as described in textbooks such as Ref.~\cite{Hirth1982}, the ^^ ^^ stacking fault octahedra" are not.
It is speculated that an uncommon defect structure, which might occur in a transient stage, was discovered.
To follow the dynamical behavior of such a defect structure in order to judge this speculation is a subject of a future research.
In a recent paper \cite{Mori2014}, along with animations, we have followed evolution of the numbers of fcc, hcp, and other type of crystalline particles in processing of erasing defect structures in Monte Carlo simulations.
Such analysis is also valid for processes of defect structure formation.
Classification of crystalline types is made on the basis of the order parameters.
Not only the total numbers of three types of crystalline particles but also spatial distribution of order parameters or types gives more detailed information.
For example, stacking faults are identified as successive hcp planes in an fcc matrix.
While geometrical crystallographic observation for dynamical processes is time-consuming, following evolution of distribution of classified particles is relatively easy.

\subsection{\label{sec:template}Periodicity of template}

Let us mention about the lattice periodicity of the template.
The periodicity of the present template is $1.0455\sigma$, which is slightly smaller than the separation of \{110\} in the flat bottom wall case \cite{Mori2006STAM}.
The reason was as follows: while in the flat wall cases~\cite{Mori2006JCP,Mori2006STAM} we concerned the effect of gravity to reduce the stacking disorders, which occurred at $g^*$ $\cong$ 0.9, in the present study the crystal at $g^*$ = 1.6 has been analyzed.
The strength of the gravity for the present study was greater than in previous studies and then the crystal was more compressed.
In the stress state produced by the combination of this template and this $g^*$ value, triangular-shaped defect structures may often arise.
In an experimental study, instead of the triangular-shaped defect structures, parallelepiped defect structures were observed \cite{Schall2004}.
The present periodicity of the template might slightly be large for the parallelepiped defect structures to arise easily at $g^*$ = 1.6.
We note that, in a recent study~\cite{Mori2014}, along with the triangular-shaped defect structures, we have also observed a parallelepiped defect structure on the same template.
Recently, Hilhorst~\textit{et~al.} have successfully introduced intentional defects at desired positions in a colloidal crystal by using templates including intentional defects \cite{Hilhorst2013}.
If the type of defect structure can be controlled by adjusting the lattice periodicity of template, a possibility to develop another defect engineering arises.

\section{\label{sec:conclusion}Concluding remarks}

We recently performed Monte Carlo simulations of the hard-sphere crystal grown on a square pattern under gravity \cite{Mori2012WJE}.
In those simulations we observed triangular-shaped defect structures.
In this paper, we have taken close looks at the snapshots.
Thereby, among several triangular-shaped defect structures, compact ones have been identified with the stacking fault tetrahedra.
We have also identified isolated vacancies, vacancy-interstitial pairs, and ^^ ^^ stacking fault octahedra."
Whereas the vacancy-interstitial pairs appeared in the upper region, the isolated vacancies were induced in the lower region.
 
We have noticed the previous work on the free energy of point defects \cite{Bennett1971,Pronk2001}.
Comparison of the present results with the predicted densities of the point defects in those papers remains as a subject of a future research.
To this end, full equilibration is necessary; we plan parallel tempering simulations using various $g^*$ values.
If equilibrium nature at each altitude in gravitaional fields holds as reported by Kanai~\textit{et~al.} \cite{Kanai2005}, direct comparisons with Refs.~\cite{Bennett1971,Pronk2001} are possible.
 
Apart from the equilibrium properties, we have discussed on the single stacking faults terminated by a Shockley partial dislocation, which is speculated to disappear in an early period.
To observe disappearance of the single staking faults is one of subjects of a future research.
Also, observation of the transformation of a stacking fault tetrahedron into an octahedron is the other one.
The octahedron is also a new structure of defects, which may appear in a transient state during the transformation.
Investigation of effect of 3D defect structures on the crystallization of the hard spheres is also a future subject. 

As already mentioned, we wish to postpone dynamical observations to confirm stability of defects structures.
Elastic theoretical calculations \cite{Mori2009,Mori2010} gave a support to the conjecture that a single stacking fault was glissile.
Such calculations will also provide an additional evidence of stability of the defect structures, in addition to the geometrical consideration.
However, elastic theoretical calculations such as done in Refs.~\cite{Mori2009,Schall2004,Hilhorst2013,Hilhorst2011} were not satisfactory.
There are three elastic fields in the colloidal epitaxy; one from the template, one due to gravity, and one yielded by defect structures.
In a previous study \cite{Mori2010}, we treated the effect of template as coherent growth, that is, a constraint that the horizontal lattice parameters were fixed, and, in addition to the self-energies of last two contributions, the cross-coupling term between them was calculated for a system including an intrinsic stacking fault.
Such calculations should be done for other defect structures in a future.

\section*{\label{sec:acknoledgement}Acknowledgment}
The authors thank Professor M.~Sato for discussions and reading the manuscript.

\bibliographystyle{model1-num-names}
\bibliography{SFT}

\newpage

\newpage

\newpage

\newpage

\newpage

\end{document}